\documentclass[runningheads]{llncs}
\usepackage[T1]{fontenc}
\usepackage{graphicx}
\usepackage{todonotes}
\usepackage{float}
\usepackage[misc,geometry]{ifsym}
\begin{document}
\title{Improving the Scan-rescan Precision of AI-based CMR Biomarker Estimation}
\titlerunning{Improving the Scan-rescan Precision of AI-based CMR Biomarkers}
%
\author{Dewmini Hasara Wickremasinghe \inst{1}\textsuperscript{(\Letter)}\orcidID{0000-0002-9202-8873} \and
Yiyang Xu\inst{1}\orcidID{0009-0000-2866-5360} \and Esther Puyol-Ant\'on \inst{1}\orcidID{0000-0002-9789-6629}\and Paul Aljabar \inst{2} \orcidID{0000-0002-6173-9007}\and Reza Razavi \inst{1}\orcidID{0000-0003-1065-3008}\and
Andrew P. King\inst{1}\orcidID{0000-0002-9965-7015}}

\authorrunning{D. H. Wickremasinghe et al.}
%
\institute{School of Biomedical Engineering \& Imaging Sciences, King\textquotesingle s College London, London, UK \\ \email{dewmini.wickremasinghe@kcl.ac.uk} \\ \and Perspectum Ltd., Oxford, UK}
\maketitle              
\begin{abstract}
Quantification of cardiac biomarkers from cine cardiovascular magnetic resonance (CMR) data using deep learning (DL) methods offers many advantages, such as increased accuracy and faster analysis. However, only a few studies have focused on the scan-rescan precision of the biomarker estimates, which is important for reproducibility and longitudinal analysis. Here, we propose a cardiac biomarker estimation pipeline that not only focuses on achieving high segmentation accuracy but also on improving the scan-rescan precision of the computed biomarkers, namely left and right ventricular ejection fraction, and left ventricular myocardial mass. We evaluate two approaches to improve the apical-basal resolution of the segmentations used for estimating the biomarkers: one based on image interpolation and one based on segmentation interpolation. Using a database comprising scan-rescan cine CMR data acquired from 92 subjects, we compare the performance of these two methods against ground truth (GT) segmentations and DL segmentations obtained before interpolation (baseline).
The results demonstrate that both the image-based and segmentation-based interpolation methods were able to narrow Bland-Altman scan-rescan confidence intervals for all biomarkers compared to the GT and baseline performances.
Our findings highlight the importance of focusing not only on segmentation accuracy but also on the consistency of biomarkers across repeated scans, which is crucial for longitudinal analysis of cardiac function. 

\end{abstract}

%
%
%
\section{Introduction}
Cardiovascular magnetic resonance (CMR) is considered the gold-standard for the assessment of the function and morphology of the heart \cite{von2016representation}. The accurate delineation of the left ventricular (LV) and right ventricular (RV) blood pools (BP), and the LV myocardium from cine CMR data allows for the computation of cardiac biomarkers such as LV and RV ejection fraction (LVEF and RVEF) and LV myocardial mass (LVM), which are crucial parameters for the assessment of cardiac health. CMR has also shown better interstudy reproducibility of cardiac biomarkers (based on manual contouring) compared to echocardiography \cite{grothues2002comparison}.
However,
manual segmentation of cardiac structures
heavily depends on the training and experience of the clinicians, introducing both intra- and inter-observer variability
\cite{petitjean2011review,suinesiaputra2015quantification}.

With the growing availability of CMR imaging data and the technological advancements in deep learning (DL) methods, many studies have investigated the advantages of using DL models for the analysis of CMR data \cite{bai2018automated,mariscal2023artificial,ruijsink2020fully}. The integration of DL methods into clinical workflows holds the potential to significantly reduce the time required for the analysis of the images, achieve a more accurate delineation of the segmented structures, and eliminate intra- and inter- observer variability \cite{galati2022accuracy,seetharam2020role}. 
As demonstrated in \cite{wang2022ai}, commercially available DL tools are now available for automated segmentation of cine CMR data, which have shown accuracy comparable to that of clinicians.


However, although the use of DL for the analysis of CMR imaging data presents many benefits and is able to achieve a high level of consistency with clinicians in quantifying cardiac function, many studies overlook the importance of the repeatability (i.e., precision) of the estimated biomarkers. The precision of biomarker estimates is an important measure of the reliability and robustness of the developed methods and can be measured from scan-rescan data acquired from the same subjects. 
A small number of papers have reported investigations into scan-rescan repeatability of DL-based CMR segmentation and biomarker estimation.
Bhuva et al. \cite{bhuva2019multicenter}
performed a scan-rescan analysis on LV biomarkers from cine CMR and demonstrated that their fully automated convolutional neural network (CNN) was able to obtain precision comparable to that of trained junior annotators, but slightly worse than that of experts.
Davies et al. \cite{davies2022precision} also assessed scan-rescan variability in CNN-based LV biomarker estimation from cine CMR compared to human analysis. They reported that the precision of a DL model surpassed that of human annotators, obtaining slightly lower coefficients of variation.

These studies show that state-of-the-art DL-based segmentation and biomarker estimation methods have precision similar to that of human annotators. However, in many CMR imaging examinations clinicians are interested in assessing the \emph{change} in biomarkers (compared to a previous scan), rather than just the absolute value. In such situations, because the changes can be very small, very high precision is needed and arguably even the precision of a human expert annotation may be insufficient. For instance, during expert segmentation, there can be significant variability in the choice of which short axis slice to stop segmenting the myocardium at the basal/apical ends of the LV \cite{marcus1999influence,marwick2018ejection}. This motivates the case for automated tools to have \emph{better} precision than experts to enable such longitudinal analyses \cite{bhuva2019multicenter}.


In this paper we (i) assess the accuracy and precision of DL-based biomarker estimation on an external validation set of scan-rescan cine CMR data, and (ii) propose methods to improve the precision of biomarker estimation. This work, to the best of our knowledge, is the first to specifically investigate methods for minimising scan-rescan variability in DL-based cardiac biomarker estimation. By addressing this gap in the literature, our study aims to enable a more robust DL-based analysis of longitudinal CMR data.

\section{Materials and Methods}
\subsection{Scan-rescan Dataset}
\textcolor{black}{The data used in the experiments consisted of a dataset of 184 scan-rescan cine CMR acquisitions obtained from 92 healthy volunteers provided by Perspectum. Each subject underwent the CMR acquisition twice within a short period of time, usually within the same day and using the same scanner. Repositioning of the patient and replanning were carried out for each of the two scans.} These CMR scans were acquired using three scanners from Siemens Healthcare (1.5T MAGNETOM Aera, 3T MAGNETOM Vida and MAGNETOM Prisma). For each subject, short-axis (SAX), two-chamber and four-chamber long-axis (LAX) views were acquired. Ground truth (GT) segmentations of the LVBP, RVBP and LV myocardium for the SAX data were manually obtained for the end-diastole (ED) and end-systole (ES) frames. The other frames were not used in the experiments.


\subsection{Overview}

\begin{figure}[h]
    \centering
    \includegraphics[width = \textwidth]{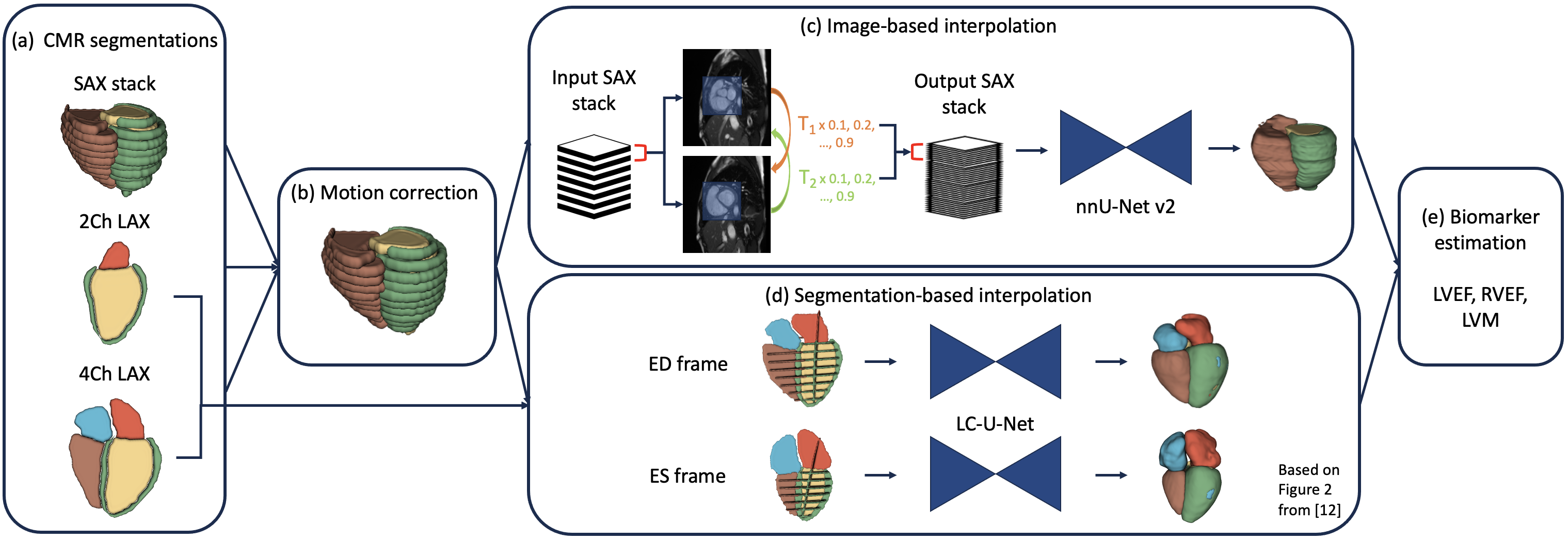}
    \caption{Overview of the motion correction and interpolation pipelines.}
    \label{fig:methods_overview}
\end{figure}

An overview of the two methods evaluated for making estimates of cardiac biomarkers with improved precision is shown in Figure \ref{fig:methods_overview}. We describe the components of the two approaches below. The image-based interpolation approach (box (c)) is a novel pipeline that we introduce here. The segmentation-based interpolation approach (box (d)) is based on the work described in \cite{muffoletto2024evaluation} but is evaluated here for the first time on scan-rescan data.

\subsection{Segmentation Models}
The segmentations of the LVBP, RVBP, and LV myocardium for the SAX data were obtained using a 2D nnU-Net v2 model \cite{isensee2021nnu}. The model was trained on 4872 SAX cine CMR scans and manual annotations from the UK Biobank \cite{sudlow2015uk}. Two more 2D nnU-Net v2 models trained on a subset of 200 cases from the UK Biobank data were used to obtain segmentations of the LVBP, left atrium (LA), and LV myocardium from the two-chamber long-axis view and the LVBP, RVBP, LA, right atrium (RA), and LV myocardium from the four-chamber long-axis view, respectively. The SAX segmentations were used for biomarker estimation for all methods whereas both the SAX and LAX segmentations were used for motion correction (box (a) in Figure \ref{fig:methods_overview}) as well as in the segmentation interpolation method (box (d)).

\subsection{Motion Correction}
A motion correction algorithm \cite{Yiyang} was applied to correct the misalignments between slices in the SAX stack due to differing breath-hold positions. This algorithm aimed to find the in-plane translations that maximised the intersection between the DL segmentations of the SAX stack data with those of the two-chamber and four-chamber long-axis data. The translations obtained from this registration process were then applied to the SAX images and DL-produced segmentations, and these were used as input data for the interpolation methods described in the following section.

\subsection{Interpolation}
Our approach to improving the precision of biomarker estimates is based on increasing the resolution of SAX images or segmentations in the through-plane direction. Two different methods were investigated:  the first was an image-based interpolation of the SAX images, and the second was an interpolation of the SAX segmentations.

\subsubsection{Image-based interpolation:} In this approach, we first estimated the non-linear deformations between each pair of adjacent slices in the SAX stack. These estimates were made in two directions, base-to-apex and apex-to-base. The employed registration algorithm was a multi-level 2D B-spline free-form deformation (MFFD) algorithm \cite{rueckert1999comparison,rueckert1999nonrigid,schnabel2001generic} implemented in MIRTK. A region of interest was used for the registration, based on masks of the heart formed by the union of the DL segmentations of the LVBP, RVBP, and LV myocardium over all slices. The resulting B-spline control point displacements for each transformation were then scaled by factors in the range $[0.1, 0.2, \ldots, 0.9]$. \textcolor{black}{The scaled transforms were applied to the source images and resampled using linear interpolation.} As a result, interpolated slices were generated at distances proportional to the scaling coefficients. This process was repeated for both registration directions and the slices from each were averaged to form each interpolated image. These interpolated images were then fed to the DL SAX segmentation model to obtain segmentations of the LVBP, RVBP, and LV myocardium. LV and RV BP volumes at ED and ES (LVEDV and LVESV, RVEDV and RVESV), and LVM were derived from the segmentations, and LVEF and RVEF values were computed from the cardiac volumes. 

\subsubsection{Segmentation-based interpolation:} This method, described in \cite{muffoletto2024evaluation}, used a 3D U-Net to predict dense segmentation label maps from sparse 2D SAX and LAX segmentations. The input data for this model was a sparse representation of the heart obtained from the intersection of the SAX segmentations with the two- and four-chamber LAX segmentations.
The dense segmentation estimates produced by the model were then used to compute the same cardiac biomarkers described above. 

\subsection{Evaluation}
First, to assess accuracy, the mean error (ME) and mean absolute error (MAE) between the ground truth (GT) and predicted biomarkers were compared for the three methods: baseline (pre-interpolation), image-based interpolation, and segmentation-based interpolation.

Second, to assess precision, the scan-rescan variability of estimates of LVEF, RVEF, and LVM was assessed using Bland-Altman analysis. This was performed for the (human) GT, the DL segmentations before interpolation (baseline) and the segmentations produced using both of the interpolation techniques. Coefficient of variation between scan and rescan was computed for each cardiac biomarker. Finally, a qualitative analysis of the results was carried out by generating visual representations of the DL produced segmentations.



\section{Experiments and Results}
\subsection{Quantitative analysis}
Table \ref{tab:errors} shows the ME and MAE obtained from the comparison of LVEDV, LVESV, RVEDV, RVESV, and LVM computed from the GT and DL-produced values. These values were generated to assess the accuracy of the obtained cardiac biomarkers, and verify that the interpolated biomarkers were within acceptable ranges. The errors computed after the two interpolation methods are similar to the baseline ones, except for RVEDV and RVESV computed from the segmentations obtained from the segmentation-based interpolation, which are slightly higher. It is also noticeable how, for most biomarkers, the image-based interpolation method obtained the lowest ME and MAE.
\begin{table}[h]
\renewcommand*{\arraystretch}{1.2}
\centering
\caption{Mean error (ME) and mean absolute error (MAE) for LVEDV, LVESV, RVEDV, RVESV, and LVM computed between the ground truth (GT) and the baseline (pre-interpolation), image-based interpolation, and segmentation-based interpolation segmentations. Lowest values for each biomarker shown in bold.}
\begin{tabular}{|ccccccc}
\hline\hline
\multicolumn{7}{c}{\textbf{Accuracy}} \\ \hline\hline 
\multicolumn{1}{l|}{} & \multicolumn{2}{c|}{Baseline} & \multicolumn{2}{c|}{Image interpolation} & \multicolumn{2}{c}{Segmentation interpolation} \\ \hline
\multicolumn{1}{c|}{} & \multicolumn{1}{c|}{\textbf{ }\textbf{ }\textbf{ }ME\textbf{ }\textbf{ }\textbf{ }\textbf{ }} & \multicolumn{1}{c|}{\textbf{ }\textbf{ }\textbf{ }MAE\textbf{ }\textbf{ }\textbf{ }\textbf{ }} & \multicolumn{1}{c|}{\textbf{ }\textbf{ }\textbf{ }\textbf{ }ME\textbf{ }\textbf{ }\textbf{ }\textbf{ }\textbf{ }} & \multicolumn{1}{c|}{\textbf{ }\textbf{ }\textbf{ }MAE\textbf{ }\textbf{ }\textbf{ }\textbf{ }} & \multicolumn{1}{c|}{\textbf{ }\textbf{ }\textbf{ }ME\textbf{ }\textbf{ }\textbf{ }\textbf{ }} & \multicolumn{1}{c}{MAE} \\ \hline
\multicolumn{1}{l|}{LVEDV (ml)} & \multicolumn{1}{r|}{4.39} & \multicolumn{1}{r|}{7.48} & \multicolumn{1}{r|}{5.69} & \multicolumn{1}{r|}{8.20} & \multicolumn{1}{r|}{\textbf{2.46}} & \multicolumn{1}{r}{\textbf{6.96}} \\
\multicolumn{1}{l|}{LVESV (ml)} & \multicolumn{1}{r|}{0.99} & \multicolumn{1}{r|}{\textbf{3.94}} & \multicolumn{1}{r|}{\textbf{0.09}} & \multicolumn{1}{r|}{4.04} & \multicolumn{1}{r|}{-2.65} & \multicolumn{1}{r}{4.84} \\
\multicolumn{1}{l|}{RVEDV (ml)} & \multicolumn{1}{r|}{-8.00} & \multicolumn{1}{r|}{9.88} & \multicolumn{1}{r|}{\textbf{-3.49}} & \multicolumn{1}{r|}{\textbf{8.30}} & \multicolumn{1}{r|}{-12.19} & \multicolumn{1}{r}{13.21} \\
\multicolumn{1}{l|}{RVESV (ml)} & \multicolumn{1}{r|}{-10.91} & \multicolumn{1}{r|}{12.18} & \multicolumn{1}{r|}{\textbf{-8.67}} & \multicolumn{1}{r|}{\textbf{10.29}} & \multicolumn{1}{r|}{-17.34} & \multicolumn{1}{r}{17.79} \\
\multicolumn{1}{l|}{LVM (g)} & \multicolumn{1}{r|}{-0.66} & \multicolumn{1}{r|}{\textbf{5.48}} & \multicolumn{1}{r|}{\textbf{-0.51}} & \multicolumn{1}{r|}{5.91} & \multicolumn{1}{r|}{-2.44} & \multicolumn{1}{r}{5.93} \\  \hline
\end{tabular}%
\label{tab:errors}
\end{table}

\begin{figure}[h]
    \centering
    \includegraphics[width=\textwidth]{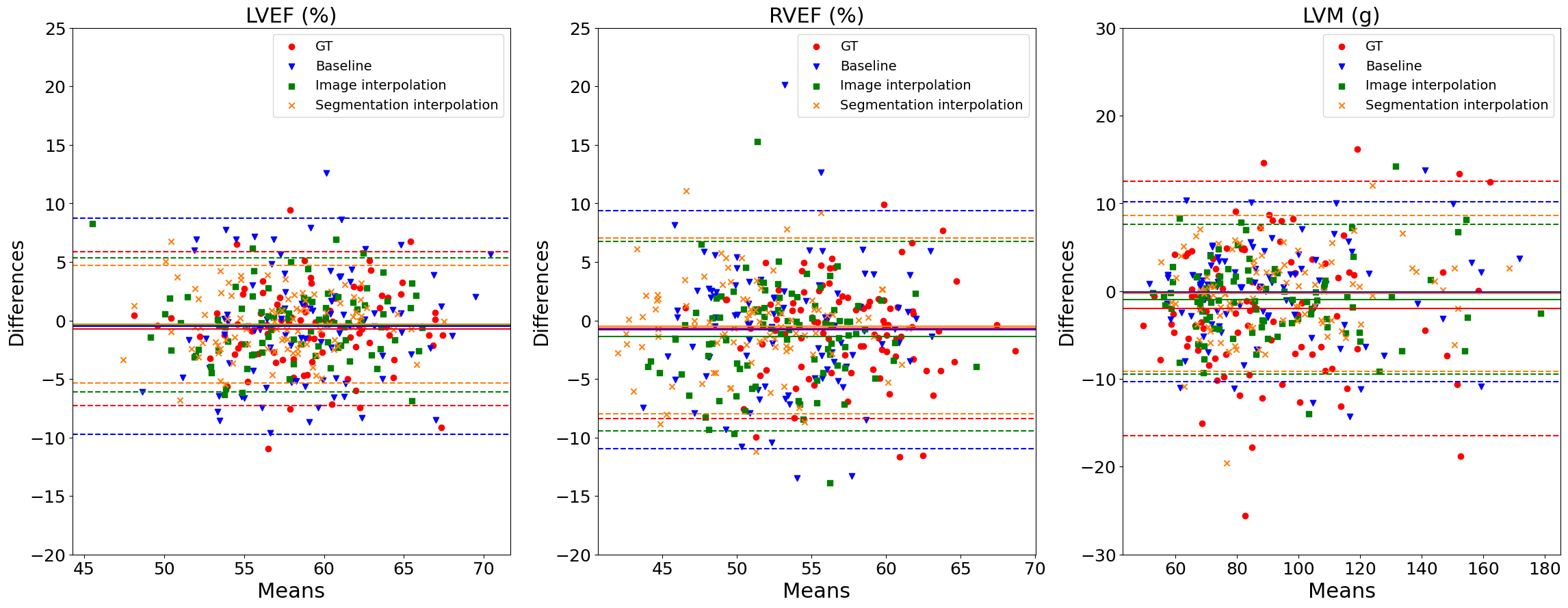}
    \caption{Bland-Altman plots for LVEF, RVEF, and LVM between scan and rescan for human ground truth (GT, red), baseline pre-interpolation (blue), image-based interpolation (green) and segmentation-based interpolation (orange).}
    \label{fig:interp_bland_altman}
\end{figure}

To quantify precision, Bland-Altman plots for each biomarker are shown in Figure \ref{fig:interp_bland_altman} for the GT and the DL-based methods, and Table \ref{tab:precision} shows the Bland-Altman confidence intervals (expressed as 
 mean $\pm$ 1.96 $\times$ standard deviation).
For LVEF and LVM, both interpolation methods showed the best agreement between scan and rescan, with narrow confidence intervals and fewer outliers compared to the GT and the baseline (pre-interpolation) approach. The confidence interval of RVEF computed from the image-based interpolation method was slightly wider than the GT one, which is likely due to the two outliers that can be seen in the plot. It is also noticeable that the baseline confidence intervals are wider than the ones for the GT for LVEF and RVEF.

\begin{table}[h]
\renewcommand*{\arraystretch}{1.2}
\centering
\caption{Bland-Altman confidence intervals and coefficient of variation values between scan and rescan for the LVEF, RVEF, and LVM computed from the ground truth (GT), baseline (pre-interpolation), image-based interpolation, and segmentation-based interpolation. Lowest values for each biomarker shown in bold.}
\begin{tabular}{lrrrr}
\hline\hline
\multicolumn{5}{c}{\textbf{Precision}} \\ \hline\hline 
\multicolumn{5}{c}{Bland-Altman Confidence Intervals} \\ \hline
\multicolumn{1}{l|}{} & \multicolumn{1}{c|}{\textbf{ }\textbf{ }\textbf{ }GT\textbf{ }\textbf{ }\textbf{ }\textbf{ }} & \multicolumn{1}{c|}{\textbf{ }\textbf{ }\textbf{ }Baseline\textbf{ }\textbf{ }\textbf{ }\textbf{ }} & \multicolumn{1}{c|}{Image interpolation} & \multicolumn{1}{c}{Segmentation interpolation} \\ \hline
\multicolumn{1}{l|}{LVEF (\%)} & \multicolumn{1}{r|}{-0.71 $\pm$ 6.57} & \multicolumn{1}{r|}{-0.49 $\pm$ 9.23} & \multicolumn{1}{r|}{-0.38 $\pm$ 5.71} & \multicolumn{1}{r}{\textbf{-0.31 $\pm$ 5.03}} \\
\multicolumn{1}{l|}{RVEF (\%)} & \multicolumn{1}{r|}{-0.66 $\pm$ 7.73} & \multicolumn{1}{r|}{-0.79 $\pm$ 10.17} & \multicolumn{1}{r|}{-1.35 $\pm$ 8.12} & \multicolumn{1}{r}{\textbf{-0.48 $\pm$ 7.50}} \\
\multicolumn{1}{l|}{LVM (g)} & \multicolumn{1}{r|}{-2.00 $\pm$ 14.48} & \multicolumn{1}{r|}{-0.06 $\pm$ 10.26} & \multicolumn{1}{r|}{\textbf{-0.94 $\pm$ 8.54}} & \multicolumn{1}{r}{-0.27 $\pm$ 8.90} \\ \hline

\multicolumn{5}{c}{Coefficient of variation} \\ \hline
\multicolumn{1}{l|}{} & \multicolumn{1}{c|}{\textbf{ }\textbf{ }\textbf{ }GT\textbf{ }\textbf{ }\textbf{ }\textbf{ }} & \multicolumn{1}{c|}{\textbf{ }\textbf{ }\textbf{ }Baseline\textbf{ }\textbf{ }\textbf{ }\textbf{ }} & \multicolumn{1}{c|}{Image interpolation} & \multicolumn{1}{c}{Segmentation interpolation} \\ \hline
\multicolumn{1}{l|}{LVEF (\%)} & \multicolumn{1}{r|}{2.10} & \multicolumn{1}{r|}{3.25} & \multicolumn{1}{r|}{2.01} & \multicolumn{1}{r}{\textbf{1.88}} \\ 
\multicolumn{1}{l|}{RVEF (\%)} & \multicolumn{1}{r|}{\textbf{2.66}} & \multicolumn{1}{r|}{3.73} & \multicolumn{1}{r|}{3.09} & \multicolumn{1}{r}{2.86} \\ 
\multicolumn{1}{l|}{LVM (\%)} & \multicolumn{1}{r|}{3.38} & \multicolumn{1}{r|}{2.22} & \multicolumn{1}{r|}{\textbf{1.91}} & \multicolumn{1}{r}{2.01} \\ \hline 
\end{tabular}%
\label{tab:precision}
\end{table}

Table \ref{tab:precision} also shows the coefficient of variation between scan and rescan for LVEF, RVEF, and LVM computed from GT, baseline (pre-interpolation), and the two interpolation methods. Both interpolation methods obtained a lower coefficient of variation for LVEF and LVM. For RVEF the best coefficient of variation was obtained for the GT segmentations, although the interpolation methods obtained similar values to the GT and improved on the performance of the baseline approach.

\subsection{Qualitative analysis}
\begin{figure}[H]
    \centering
    \includegraphics[width=\textwidth]{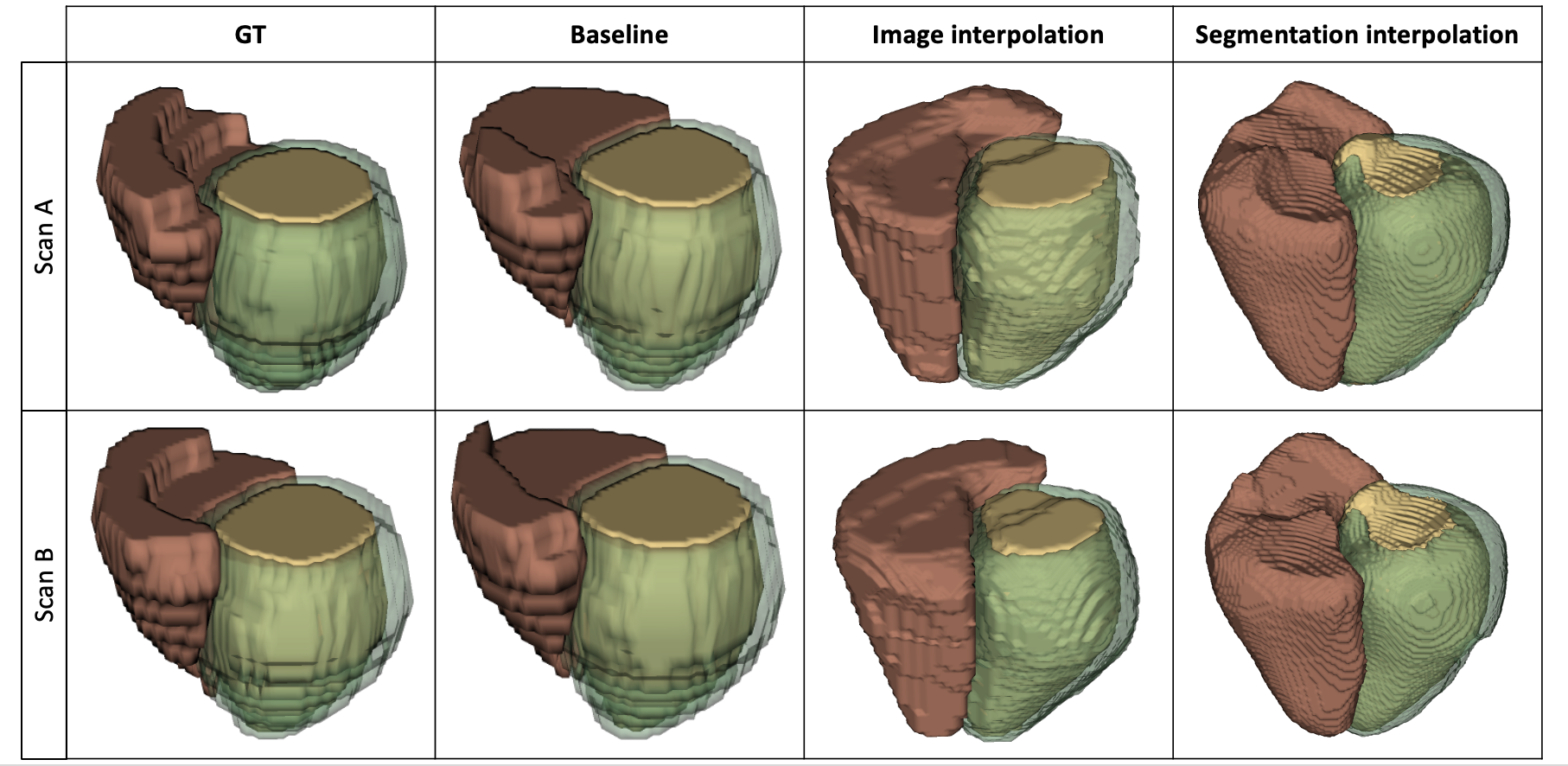}
    \caption{\textcolor{black}{Visual representations of the ground truth (GT), baseline (pre-interpolation), image-based interpolation, and segmentation-based interpolation segmentations.}}
    \label{fig:visual_res}
\end{figure}

Figure \ref{fig:visual_res} shows visual examples of the GT and DL-derived segmentations of the LVBP, RVBP, and LV myocardium for sample scan and rescan images. Both interpolation approaches produce smoother and more realistic looking segmentations.

\section{Discussion and Conclusion}
This study has focused on using interpolation methods to estimate cardiac biomarkers from cine CMR data, with the aim to improve scan-rescan precision. Two different methods were assessed to evaluate the effect of through-plane interpolation of the SAX data on the repeatability of biomarker estimation. The first one was an image-based interpolation approach, and the second was a segmentation-based interpolation method. All experiments were performed on a database of scan-rescan cine CMR data from 92 subjects.

First, to assess the accuracy of the segmentations obtained after applying the two interpolation methods, the ME and MAE were computed between the GT cardiac volumes at ED and ES, and the DL-derived ones. Due to the different granularity (i.e. through-plane resolution) of the original SAX stack (and hence the GT segmentations) and the interpolated data, direct comparisons of error values between lower resolution and higher resolution data should be treated with caution. However, such a comparison does serve as a general indicator of the accuracy of the computed biomarkers. As seen in Table \ref{tab:errors}, both post-interpolation segmentations obtained error values similar to the ones obtained from the baseline approach, with the image-based interpolation method showing accuracy higher than the baseline for most of the cardiac biomarkers. 

Scan-rescan precision was evaluated through Bland-Altman confidence limits and coefficient of variation. As summarised in Figure \ref{fig:interp_bland_altman} and Table \ref{tab:precision}, both interpolation methods showed good agreement between scan and rescan for LVEF, RVEF, and LVM. For LVEF and RVEF, the GT scan-rescan agreement was higher than the baseline, whereas the use of interpolation, whether image- or segmentation-based, always improved the agreement of scan-rescan biomarkers. This outcome not only shows the potential of the presented methods to enhance the precision of biomarker estimation, but also highlights the need for methods that focus on improving agreement between scans of the same subject, rather than solely prioritising accuracy against GT segmentations. 

In line with the results shown from the Bland-Altman analysis, the coefficient of variation values in Table \ref{tab:precision} indicate that both interpolation approaches achieved better precision between scan and rescan biomarkers compared to the baseline approach and, in most cases, were also able to improve on the GT precision. 

Other methods that focus on increasing the through-plane resolution of the SAX stack have been proposed  \cite{chen2024motion,stolt2023nisf,xia2021super} but these have not yet been evaluated for their impact on scan-rescan precision.
\textcolor{black}{Future work will focus on improving the proposed image-based interpolation method by including LAX data to enable better identification of the basal plane of the heart, which can be used as cut off when segmenting the SAX stack, and on evaluating other learning-based image interpolation approaches, such as the ones mentioned above.} However, we believe that the results presented here serve as a valuable proof of concept, demonstrating that interpolation can enhance the agreement in scan-rescan biomarkers, bringing us closer to reliable longitudinal analysis of changes in cardiac function.


\begin{credits}
\subsubsection{\ackname} \textcolor{black}{We would like to acknowledge funding from the EPSRC Centre for Doctoral Training in Medical Imaging (EP/S022104/1) and the Wellcome/EPSRC Centre for Medical Engineering at King’s College London (WT 203148/Z/16/Z). We also acknowledge funding from Perspectum Ltd., Oxford, UK.}

\subsubsection{\discintname}
The authors have no competing interests to declare that are relevant to the content of this article.
\end{credits}

%
%
%
\bibliographystyle{splncs04}
\bibliography{paper}
\end{document}